\documentclass[10pt,a4paper,onecolumn]{elsarticle}

\usepackage{epsfig}

\begin{document}

\begin{frontmatter}

\title{The Main Results of the Borexino Experiment}

\author{\small{A Derbin$^f$ and V Muratova$^f$ on behalf of the Borexino collaboration:\\
M Agostini$^{n}$, K Altenm\"{u}ller$^n$, S Appel$^n$, G Bellini$^h$, J Benziger$^k$, D Bick$^s$, G Bonfini$^e$, D
Bravo$^q$
% ...
}}

\address{
a) AstroParticule et Cosmologie, Universit\'e Paris Diderot, CNRS/IN2P3, CEA/IRFU, Observatoire de Paris, Sorbonne Paris Cit\'e, 75205 Paris Cedex 13, France\\
b) Joint Institute for Nuclear Research, Dubna 141980, Russia\\
c) Dipartimento di Fisica, Universit\`a degli Studi e INFN, Genova 16146, Italy\\
d) M. Smoluchowski Institute of Physics, Jagiellonian University, Crakow, 30059, Poland\\
e) INFN Laboratori Nazionali del Gran Sasso, Assergi 67010, Italy\\
f) St. Petersburg Nuclear Physics Institute NRC Kurchatov Institute, Gatchina 188350, Russia\\
g) NRC Kurchatov Institute, Moscow 123182, Russia\\
h) Dipartimento di Fisica, Universit\`{a} degli Studi e INFN, Milano 20133, Italy\\
i) Max-Planck-Institut f\"ur Kernphysik, 69117 Heidelberg, Germany\\
j) Dipartimento di Chimica, Universit\`{a} e INFN, Perugia 06123, Italy\\
k) Chemical Engineering Department, Princeton University, Princeton, NJ 08544, USA\\
l) Physics Department, Princeton University,Princeton, NJ 08544, USA\\
m) Gran Sasso Science Institute (INFN), 67100 £'Aquila, Italy\\
n) Physik Department, Technische Universit\"at M\"unchen, Garching 85747, Germany\\
o) Institute for Nuclear Research, Kiev 03680, Ukraine\\
p) Physics Department, University of Massachusetts, Amherst MA 01003, USA\\
q) Physics Department, Virginia Polytechnic Institute and State University, Blacksburg, VA 24061, USA\\
r) Lomonosov Moscow State University Skobeltsyn Institute of  Nuclear Physics, Moscow 119234, Russia\\
s) Institut f\"ur Experimentalphysik, Universit\"at, 22761 Hamburg, Germany\\
t) Department of Physics, University of Houston, Houston, TX 77204, USA\\
u) Physics and Astronomy Department, University of California Los Angeles (UCLA), Los Angeles, CA 90095, USA\\
v) Department of Physics, Technische Universitat Dresden, 01062 Dresden, Germany\\
w) Dipartimento di Fisica e Scienze della Terra Università degli Studi di Ferrara e INFN, 44122 Ferrara, Italy\\
x) Amherst Center for Fundamental Interactions and Physics Department, University of Massachusetts, Amherst, Massachusetts 01003,
USA\\
y) Institute of Physics and Excellence Cluster PRISMA, Johannes Gutenberg-Universit$\ddot{a}$t Mainz, 55099 Mainz, Germany\\
z) National Research Nuclear University MEPhI (Moscow Engineering Physics Institute), 115409 Moscow, Russia\\
.\\
 \bf{Proceedings of the Third Annual Large Hadron Collider Physics Conference, \\St. Petersburg, Russia, 2015}}



\begin{abstract}
The main physical results on the registration of solar neutrinos and the search for rare processes obtained by the
Borexino collaboration to date are presented.
\end{abstract}

\end{frontmatter}
\section{INTRODUCTION}

The study of solar neutrinos is at the intersection of elementary particle physics and astrophysics. On one hand these
neutrinos allow for the study of neutrino oscillations, and on the other they provide key information for accurate
solar modeling. The Borexino first detected and then precisely measured the flux of the $^7\rm{Be}$ solar neutrinos,
ruled out any significant day-night asymmetry of their interaction rate, performed the measurement of
$^8\rm{B}$-neutrino with 3 MeV threshold, made the first direct observation of the pep neutrinos, and set the tightest
upper limit on the flux of solar neutrinos produced in the CNO cycle.

The uniquely low background level of the Borexino detector made it possible to set new limits on the effective magnetic
moment of the neutrino, on the stability of the electron for decay into a neutrino and a photon, on the heavy sterile
neutrino mixing in $^8\rm{B}$-decay, on the possible violation of the Pauli exclusion principle, on the flux of high
energy solar axions and on some other rare processes.



\section{The Borexino detector}

Borexino is a real-time liquid scintillator detector for solar neutrino spectroscopy located at the Gran Sasso
Underground Laboratory \cite{Bel_2014}. Its main goal is to measure low-energy solar neutrinos via $(\nu,e)$-scattering
in an ultrapure liquid scintillator. At the same time, the extremely high radiopurity of the detector and its large
mass allow it to be used for the study other fundamental questions in particle physics and astrophysics.
\begin{figure}[h]
 \centerline{\includegraphics[bb = 60 460 500 755, width=11cm,height=4.5cm]{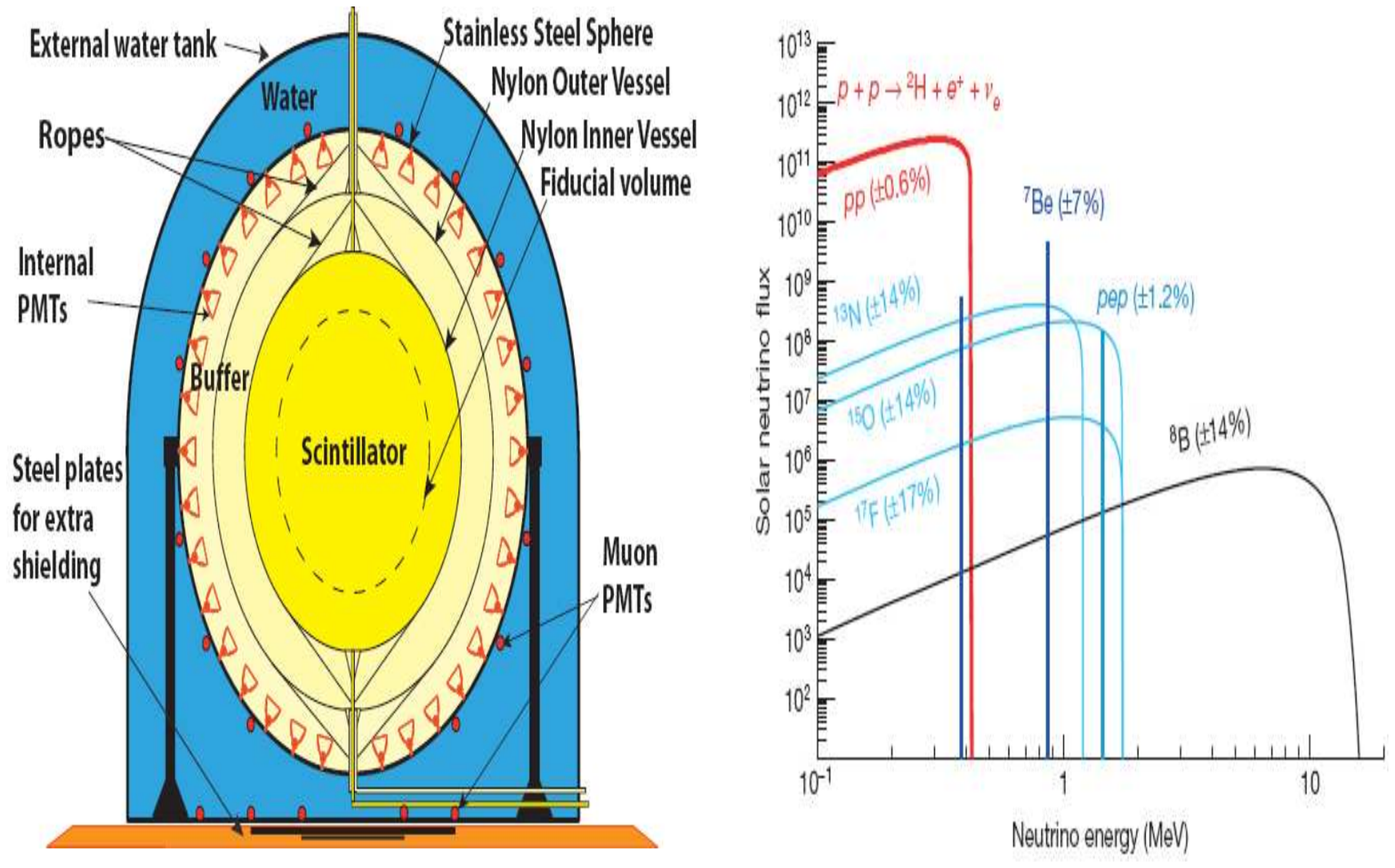}}
  \caption{Left: the schematic view of the Borexino detector. Right: solar neutrino energy spectrum predicted by standard solar model.}
\end{figure}

The detector energy and spatial resolution were studied with radioactive sources placed at different positions inside the inner
vessel. For high energies the calibration was performed with a AmBe neutron source \cite{Bac_2012}. The energy resolution scales
approximately as $\sigma_E/E = 5\%E^{-1/2}$. The position of an event is determined using a photon time of flight reconstruction
algorithm. The resolution of the event  reconstruction, as measured using the $^{214}\rm{Bi}$ - $^{214}\rm{Po}$  decay sequence,
is 13.2 cm.


The fluxes and the energy spectra of solar neutrinos from pp-chain and CNO-cycle are predicted by solar models. Thanks
to the unprecedented low background level achieved in the scintillator, Borexino already measured the fluxes and
electron recoil spectra of neutrinos coming from the $pp-$, $pep-$, $^7\rm{Be}$ -and  $^8\rm{B}$- nuclear reactions
which take place inside the Sun.

\section{$^7\rm{Be}$-neutrinos}

Borexino was designed to measure the spectrum of recoil electrons from 862 keV neutrino due to EC-process: $^7\rm{Be} +
e^- \rightarrow ^7\rm{Li} + \nu_e$. The measured count rate of  $^7\rm{Be}$-neutrino is \cite{Bel_2011}:
$\rm{R(^7Be)=46.0 \pm 1.5 (stat) \pm 1.6 (syst)}$ counts/(d 100 t). Study on a possible asymmetry between day and night
$^7\rm{Be}$-neutrino interaction rate gives \cite{Bel_2012}: $\rm{A_{dn} = 0.001 \pm 0.012 (stat) \pm 0.07 (syst)}$.
Borexino excluded the LOW region of the MSW parameter space for neutrino without the use of reactor anti-neutrino data
and therefore without the assumption of CPT symmetry.

\begin{figure}[h]
 \centerline{\includegraphics[bb = 60 500 500 765, width=11cm,height=4.5cm]{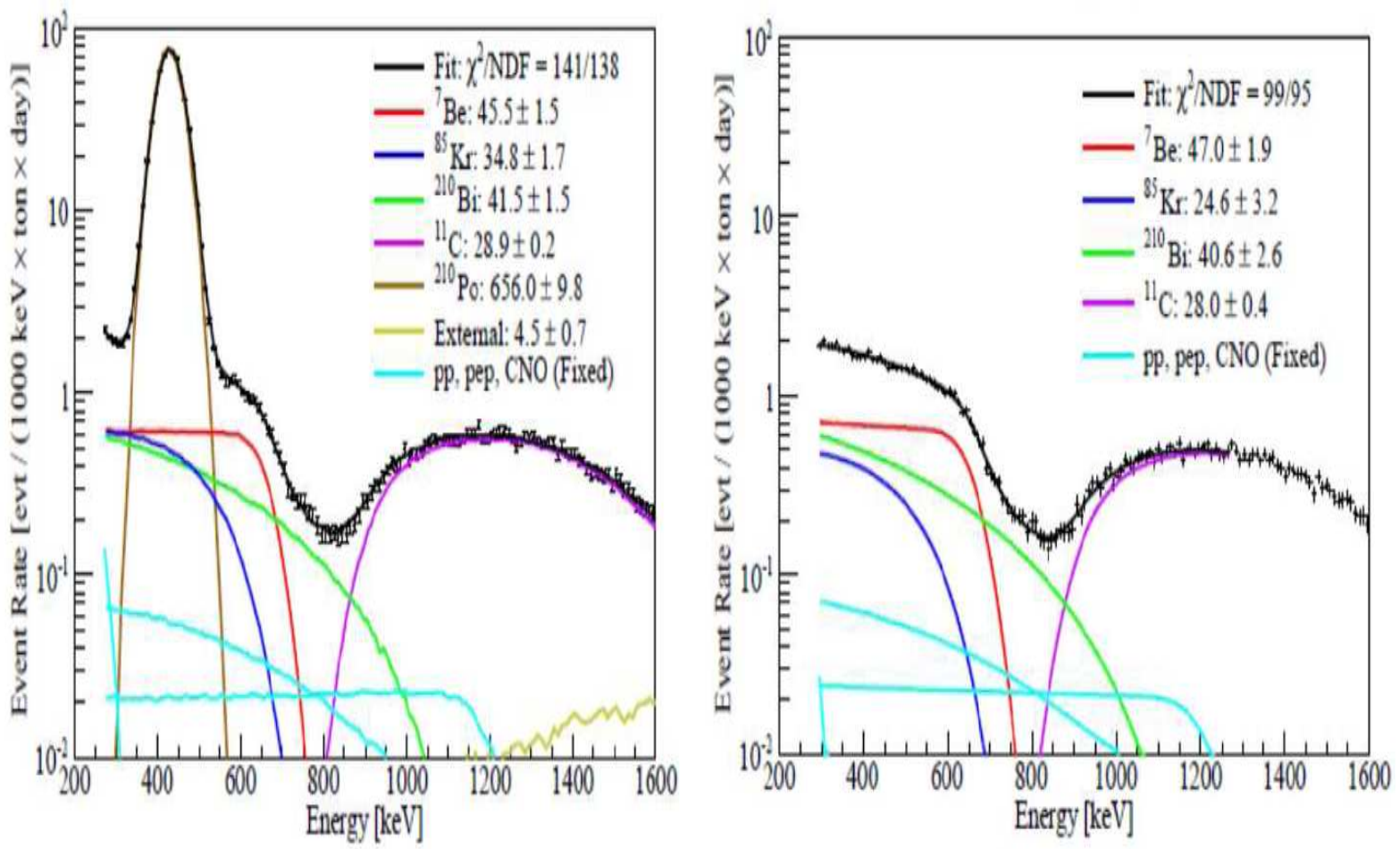}}
  \caption{Two example fitted spectra. Left: A Monte Carlo based fit over the energy region 270--1600 keV to a spectrum from
which some, but not all, of the $\alpha$ events have been removed. Right: An analytic fit over the 290--1270 keV energy
region to a spectrum obtained with $\alpha$ subtraction.}
\end{figure}

\section{$^8\rm{B}$-neutrinos}
Borexino reported the first measurement of $^8\rm{B}$ solar neutrino rate with 3 MeV threshold \cite{Bel_2010}
$\rm{R(^8B) = 0.22 \pm 0.04 (stat) \pm 0.01 (syst)}$ counts/(d 100 t) in good agreement with measurements from SNO and
SuperKamiokaNDE.

\section{pep- and CNO-neutrino}
Standard Solar Model provides an very accurate (1.2\%) flux prediction for 1.44 MeV neutrinos emitted in $p + p + e^-
\rightarrow d + \nu_e$ reaction. Borexino performed the first measurement of the $pep$-neutrino interaction rate and
set the strongest limit on the CNO neutrino interaction rate (at present, it is not sufficient to solve the High/Low
metallicity problem) \cite{Bel_2012A}: $\rm{R}(pep)=\rm{3.1 \pm 0.6 (stat) \pm 0.3(syst)}$ counts/(d 100 t) and
$\rm{R(CNO)\leq 7.9}$ counts/(d 100 t) at 95\% C.L..

\begin{figure}[h]
  \centerline{\includegraphics[bb = 60 550 500 785, width=11cm,height=4.5cm]{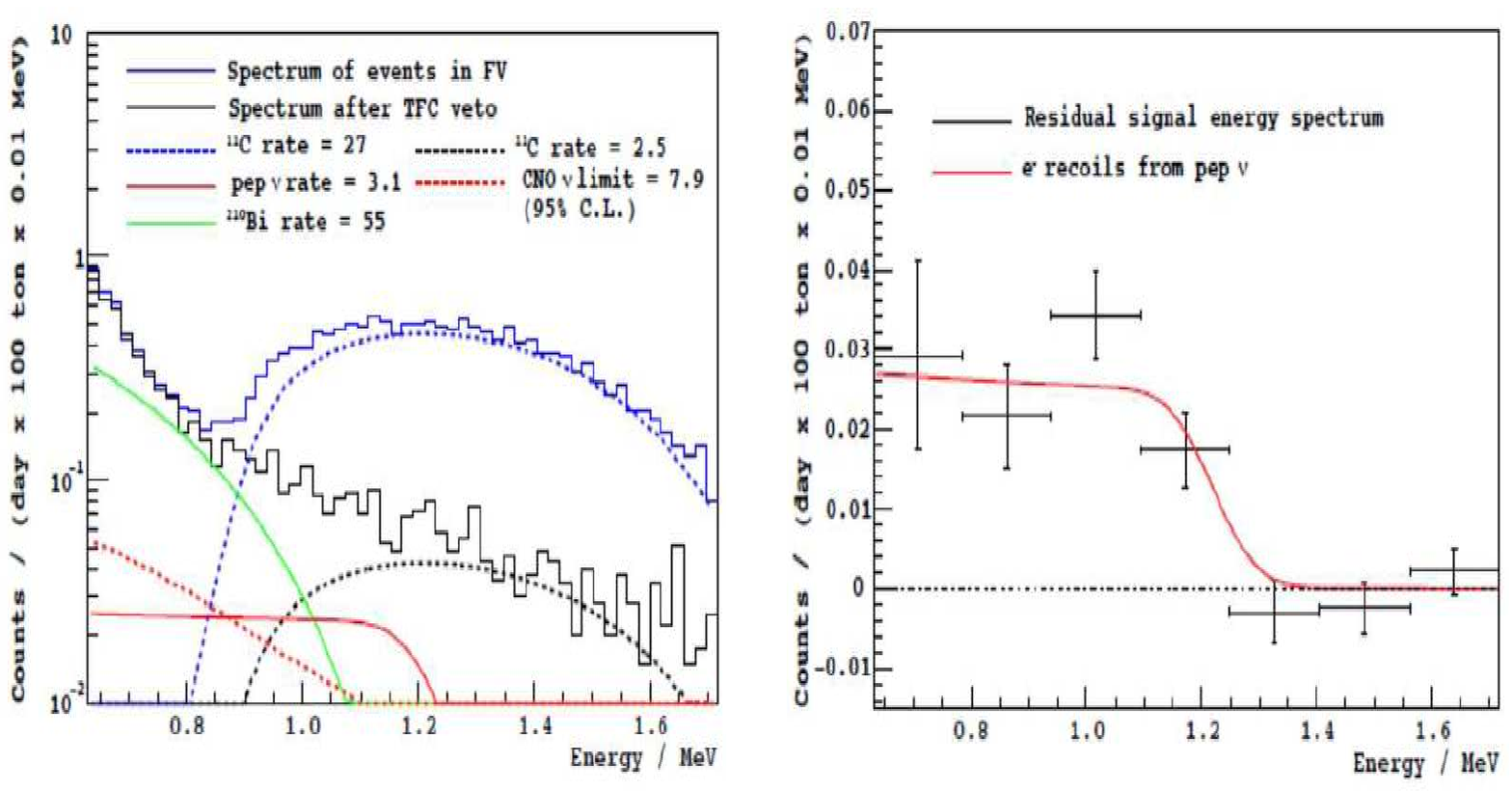}}
\caption{Left: energy spectra of the events in the fiducial volume before and after the the threefold coincidence veto
($\mu, n, ^{11}\rm{C}$ is applied. Right: residual energy spectrum after best-fit rates of all considered backgrounds
are subtracted. The e--recoil spectrum from pep-$\nu$ at the best-fit rate is shown for comparison.}
\end{figure}

\section{pp-neutrino}
Neutrino produced from the fusion of two protons for the first time has been detected in a real time detector. The
unique properties of the Borexino provided an opportunity to extract $pp$-neutrino spectrum from the background
components \cite{Bel_2014A}: $\rm{R(\it{pp})= 44 \pm 13 (stat)\pm 10 (syst)}$ counts/(d 100 t). Assuming LMA-MSW
solution this value corresponds to solar $pp$-neutrino flux $\Phi(pp) = (6.6 \pm 0.7)\times 10^{10} \rm{cm^{-2}
s^{-1}}$ which is in good agreement with the prediction of the standard solar model.
\begin{figure}[h]
  \centerline{\includegraphics[bb = 60 350 500 755, width=7cm,height=4.5cm]{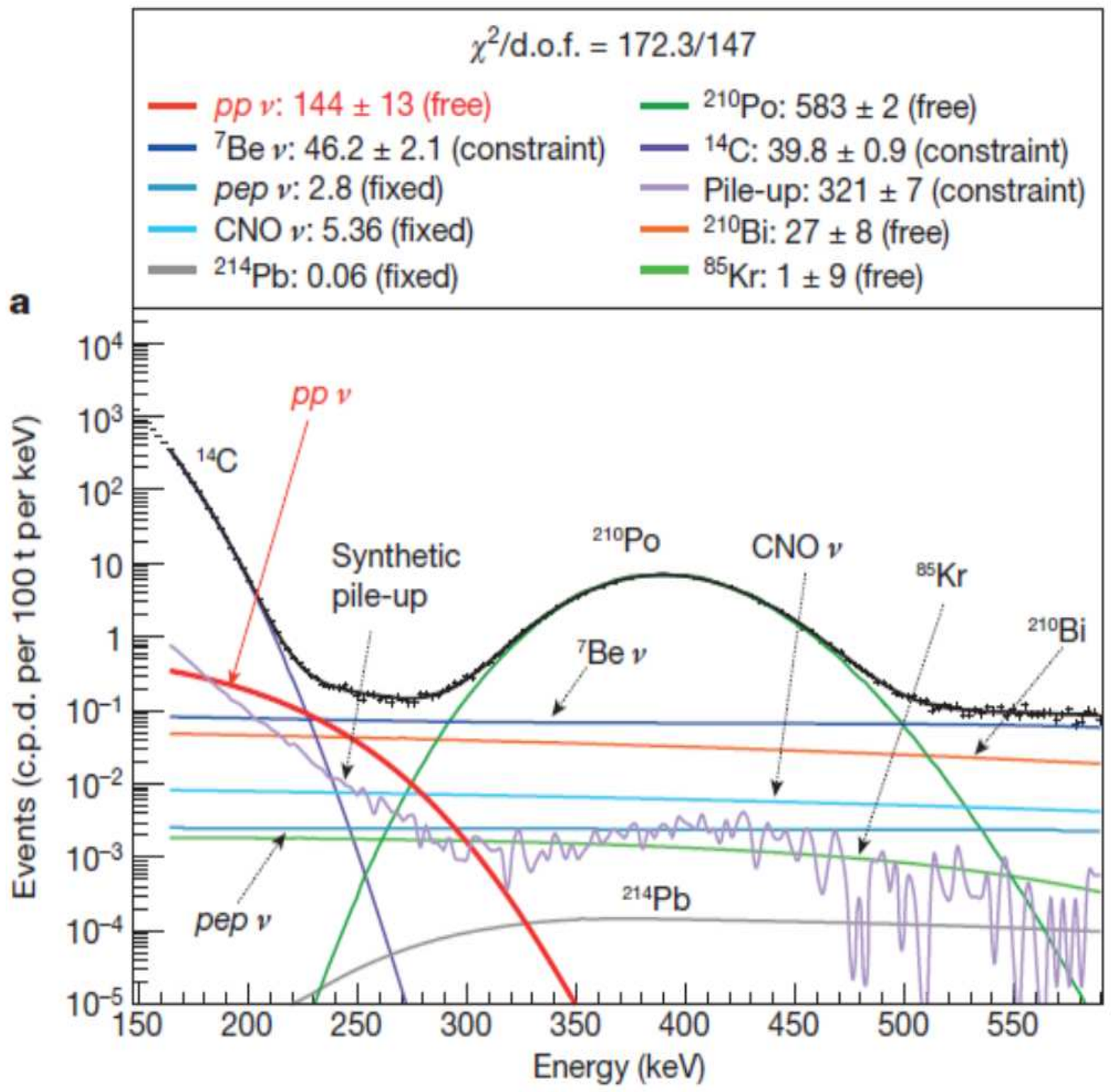}}
  \caption{Fit of the energy spectrum between 165 and 590 keV. a, The bestfit
pp neutrino component is shown in red, the $^{14}\rm{C}$ background in dark purple and the synthetic pile-up in light
purple. The large green peak is 210Po a-decays. $^7\rm{Be}$ (dark blue), $pep$- and CNO (light blue) solar neutrinos,
and $^{210}\rm{Bi}$ (orange) are almost flat in this energy region. The values of the parameters (in c.p.d. per 100 t)
are in the inset above the figure.}
\end{figure}

\section{Electron neutrino survival probability}
Survival probability of electron-neutrinos produced by the different nuclear reactions in the Sun. All the numbers are
from the Borexino. Because $pp$- and $^8\rm{B}$-neutrino are emitted with a continuum of energy the reported $P_{ee}$
value refers to the energy range contributing to the measurement. The violet band corresponds to the $\pm 1\sigma$
prediction of the MSW-LMA solution.

\begin{figure}[h]
 \centerline{\includegraphics[bb = 60 450 500 735, width=7cm,height=4.5cm]{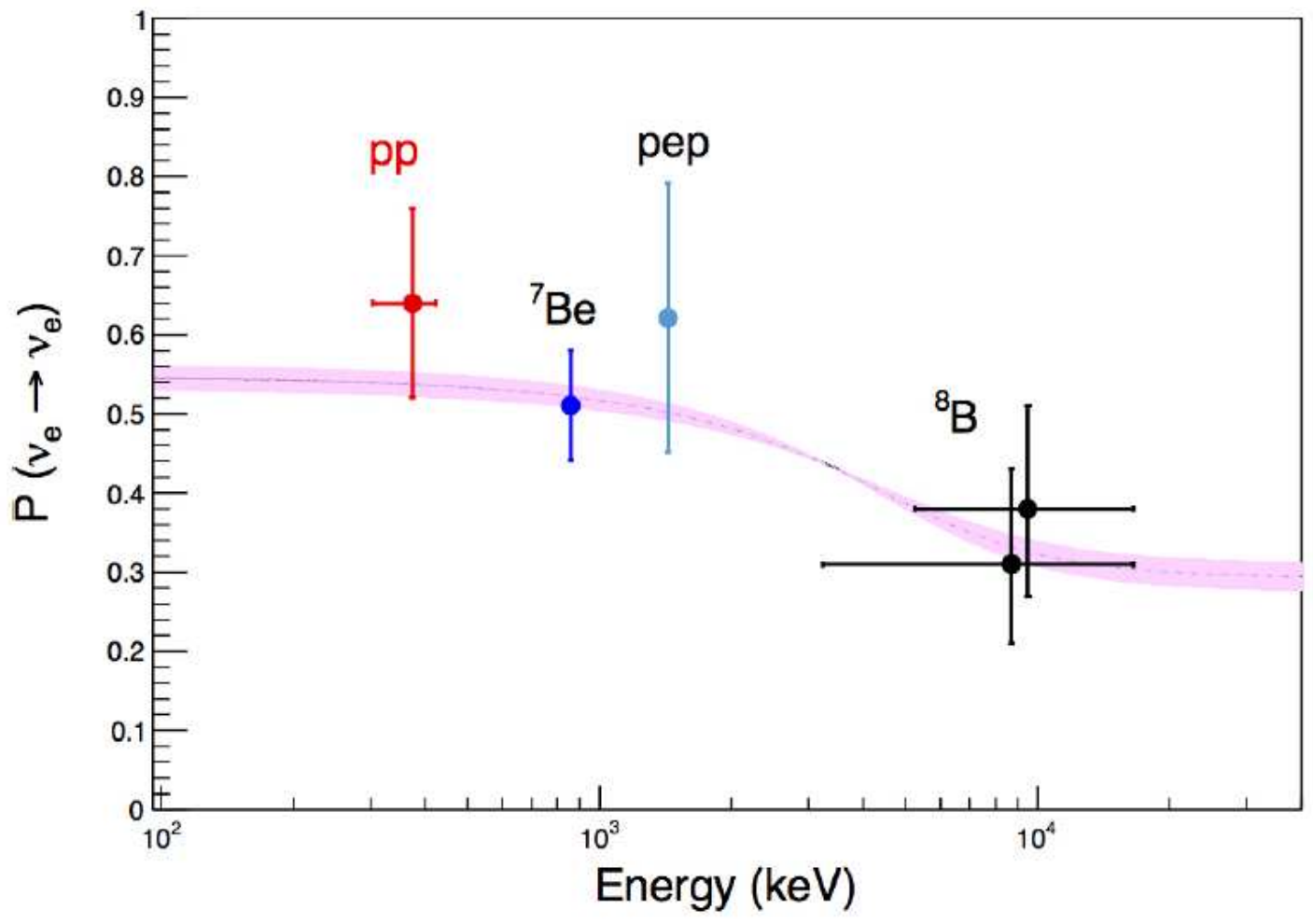}}
  \caption{Electron neutrino survival probability obtained by Borexino as a function of energy.}
\end{figure}

\section{Neutrino magnetic moment}

The shape of the electron recoil  spectrum is sensitive to the possible presence of a non-null magnetic moment, and the
sensitivity is enhanced at low energy since ~ $E_e^{-1}$. For solar neutrinos we detect the effective magnetic moment,
which is composition of magnetic moments for mass or flavor eigenstates. Borexino obtained the upper limit
\cite{Arp_2008}: $\mu_{eff} \leq 5.4\times10^{-10} \mu_B$ (90\% C.L.).

\begin{figure}[h]
 \centerline{\includegraphics[bb = 60 290 500 755, width=7cm,height=4.5cm]{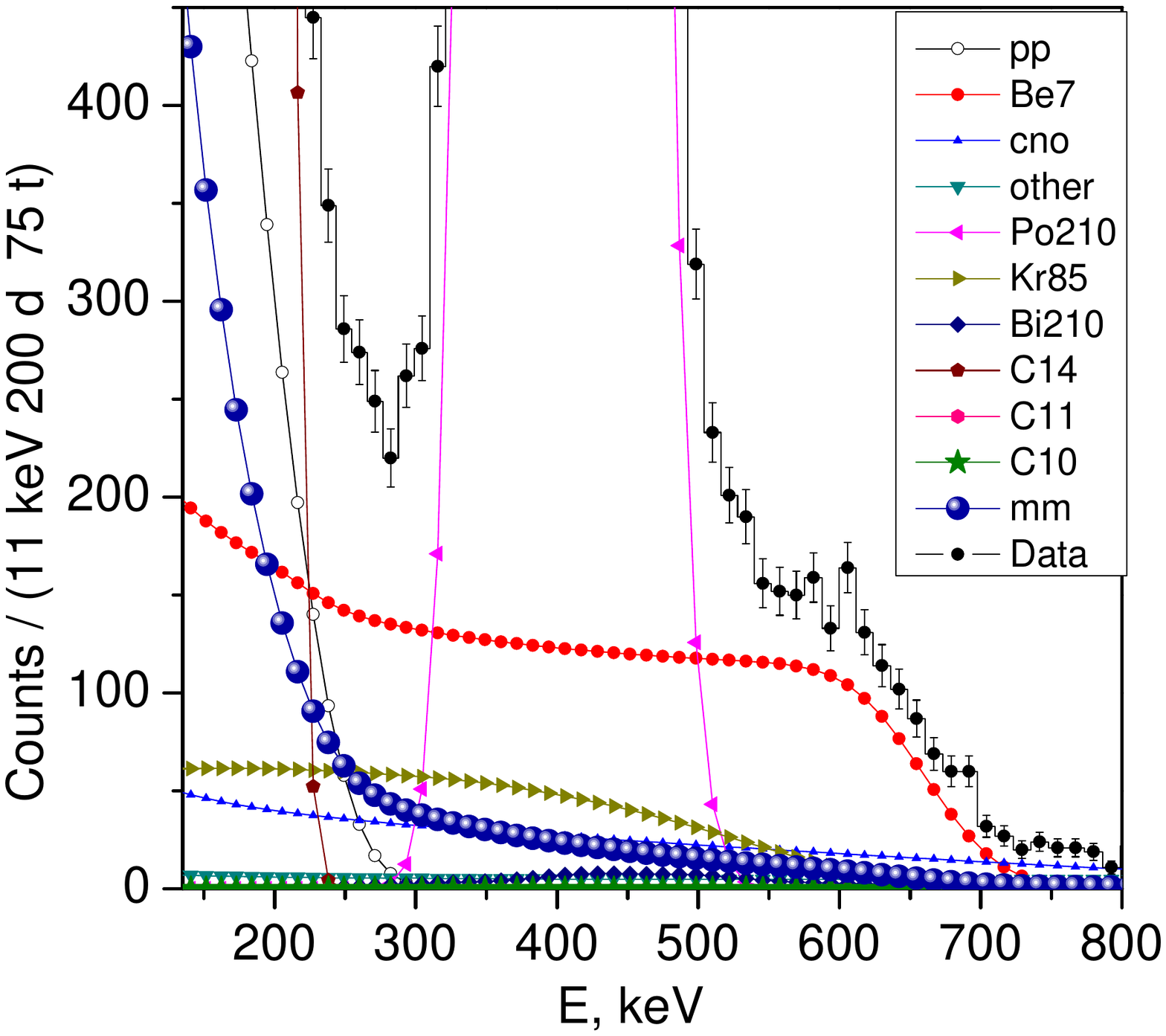}}
\caption{The e-recoil spectrum due to magnetic moment equals $5.4\times 10^{-11}\mu_B$ (blue cycles) in comparison with
others components of the Borexino data. }
\end{figure}

\section{Detection of geo- and reactor neutrinos}

Geo-neutrinos are electron anti-neutrinos produced by decays of long-lived isotopes, which are naturally present in the
interior of the Earth, such as decays in the $^{238}\rm{U}$ and $^{232}\rm{Th}$ chains, and $^{40}\rm{K}$. Results from
2056 days of data taking correspond to exposure of $(5.5 \pm 0.3)\times 10^{31}$ proton$\times$yr. Assuming a
chondritic Th/U mass ratio of 3.9, Borexino detected $(23.7^{+6.5}_{-5.7})$ geo-neutrino events and
$(52.7^{+8.5}_{-7.7})$ reactor (anti)neutrinos  \cite{Ago_2015}. The Borexino reported on the search for anti-neutrinos
of yet unknown origin and, in particular,  set a new upper limit for a hypothetical solar $ \tilde{\nu}$ flux of 760
$\rm{cm}^{-2}\rm{s}^{-1}$, obtained assuming an undistorted solar $^8\rm{B}$ energy spectrum \cite{Bel_2011A}.

\begin{figure}[h]
 \centerline{\includegraphics[bb = 60 500 500 775, width=7cm,height=4.5cm]{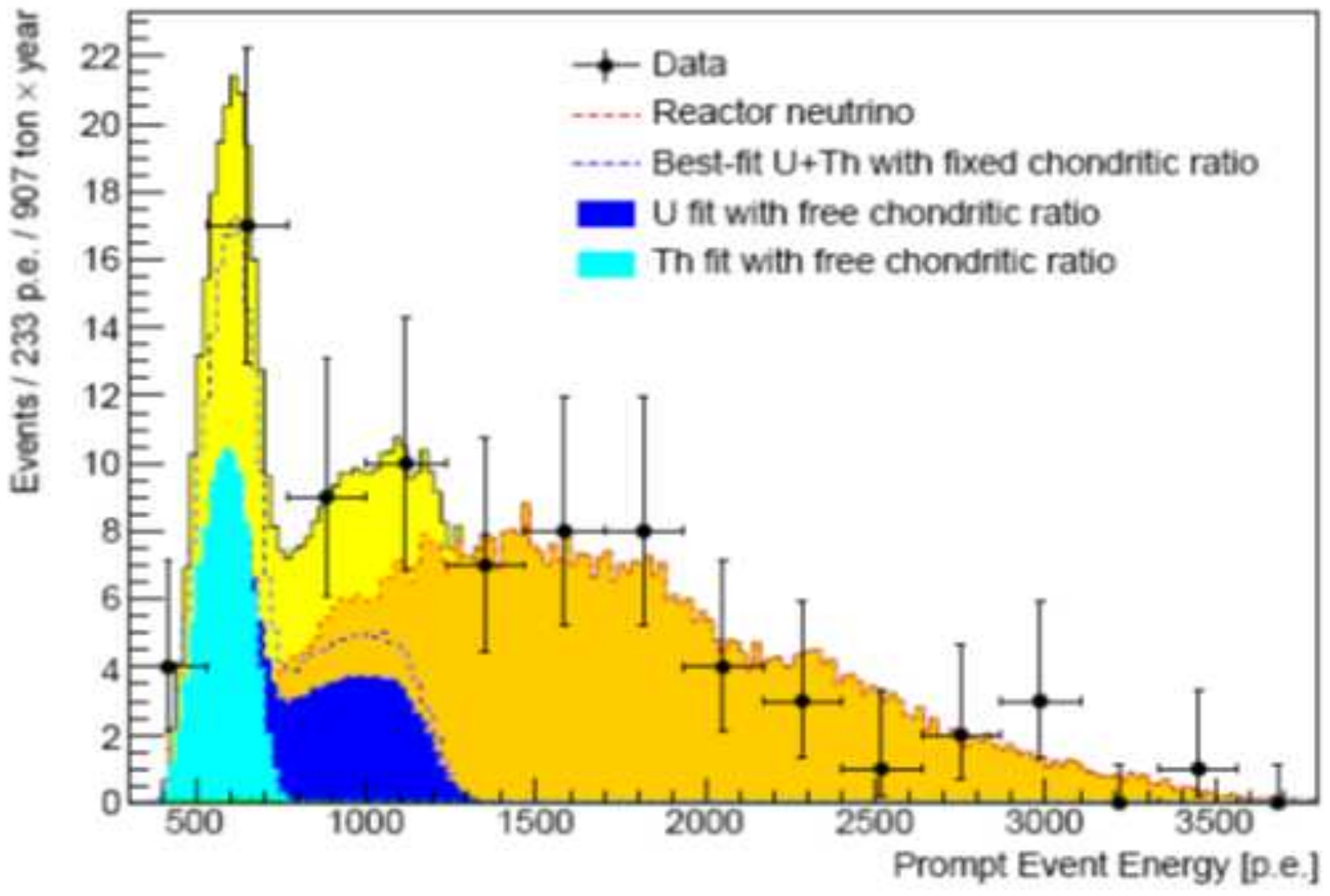}}
  \caption{Prompt light yield spectrum, in units of photoelectrons (p.e.), of $\tilde{\nu}_e$ candidates and the best-fit. The best-fit shows the geo-neutrino and reactor neutrino spectra (dotted lines)
assuming the chondritic ratio. Colored areas show the result of a separate fit with U (blue) and Th (light blue) set as free and
independent parameters.}
\end{figure}

\section{Heavy sterile neutrino}

The Borexino constrains the mixing of a heavy neutrino with mass 1.5 MeV $\leq m_H \leq$ 14 MeV appearing in
$^8\rm{B}$-decay to be $|U_{eH}|^2 \leq (10^{-3} - 4\times10^{-6})$, respectively \cite{Bel_2013}. These limits are 10
to 1000-fold stronger than those obtained by experiments searching for $\nu_H \rightarrow \nu_L + e^+ + e^-$ decays at
nuclear reactors and 1.5-4 times stronger than those inferred from $\pi \rightarrow e + \nu$ decay.

\begin{figure}[h]
 \centerline{\includegraphics[bb = 60 320 500 745, width=8cm,height=5.5cm]{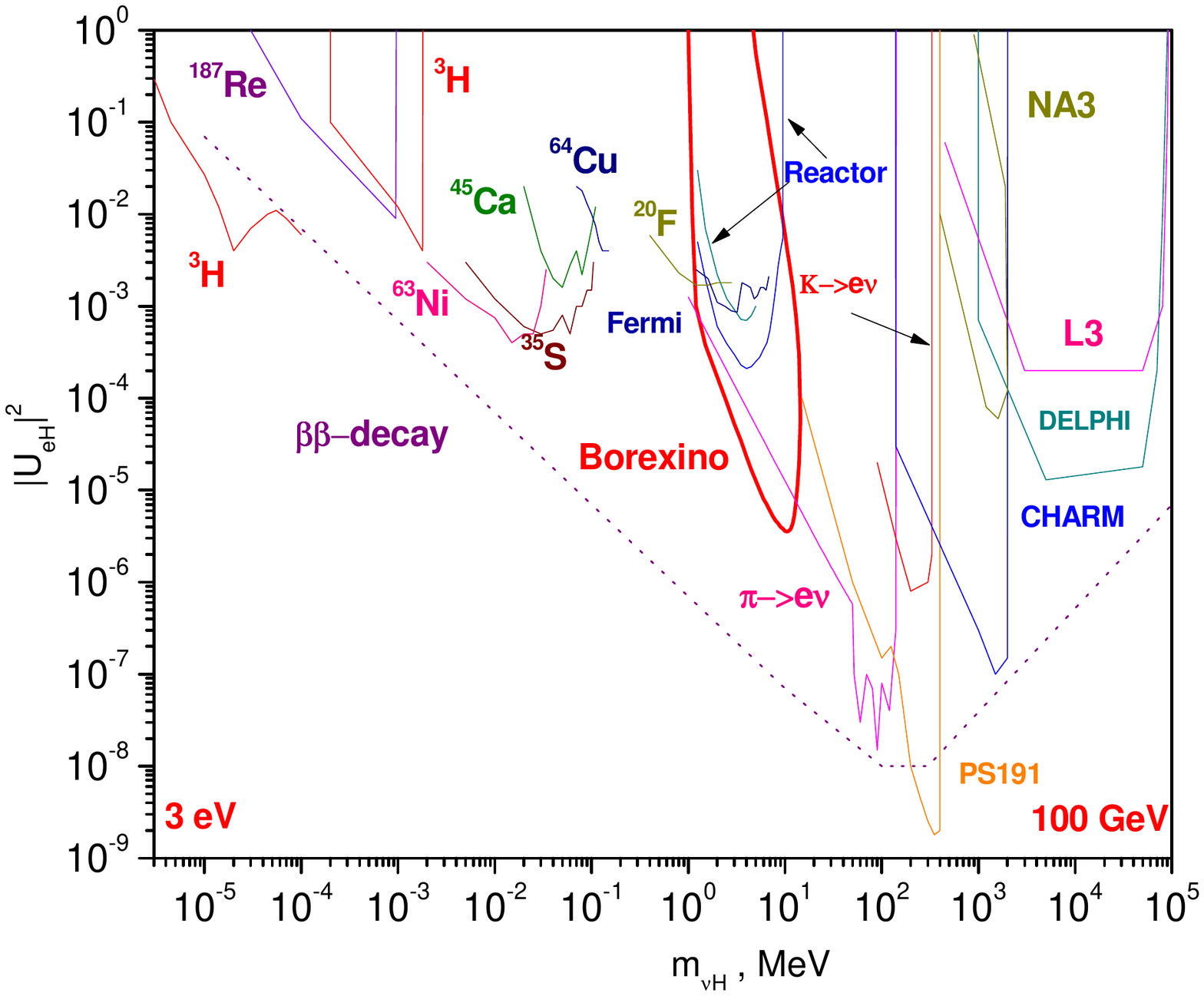}}
\caption{The Borexino constraints (red) and limits on $|U_{eH}|^2$ versus $m_H$ in the mass range (3 eV -- 100 GeV)
from different experiments.}
\end{figure}

\section{Test of Pauli Exclusion Principle}
Using the unique features of the Borexino detector the following new limits on non-paulian transitions of nucleons from
the $1P_{3/2}$-shell to the filled $1S_{1/2}$-shell in $^{12}$C with the emission of $\gamma, n, p$ and $\beta^{\pm}$
particles have been obtained \cite{Bel_2010A}: $\tau(^{12}$C$\rightarrow{^{12}\widetilde{\rm{C}}}+\gamma) \geq
5.0\times10^{31}$~y, $\tau(^{12}$C$\rightarrow{^{11}\widetilde{\rm{B}}}+ p) \geq 8.9\times10^{29}$~y,
$\tau(^{12}$C$\rightarrow{^{11}\widetilde{\rm{C}}}+ n) \geq 3.4 \times10^{30}$~y,
$\tau(^{12}$C$\rightarrow{^{12}\widetilde{\rm{N}}}+ e^- + \nu) \geq 3.1\times10^{30}$~y and
$\tau(^{12}\rm{C}\rightarrow{^{12}\widetilde{\rm{B}}}+ e^+ + \overline{\nu}) \geq 2.1\times10^{30}$ y, all with 90\%
C.L. These limits are the best to date.

\section{High energy solar axions}

A search for 5.5-MeV solar axions produced in $p + d \rightarrow ^3\rm{He + A}$ (5.5 MeV) reaction was performed
\cite{Bel_2012B}. The Compton conversion of axions to photons - $\rm{A + e^- \rightarrow e^- + \gamma}$; the axio-electric effect
- $\rm{A + e^- + Z \rightarrow e^- + Z}$; the decay of axions into two photons - $\rm{A \rightarrow 2\gamma}$; and inverse
Primakoff conversion on nuclei - $\rm{A + Z \rightarrow Z + \gamma}$, are considered. Model independent limits on axion-electron
($g_{Ae}$), axion-photon ($g_{A\gamma}$), and isovector axion-nucleon ($g^3_{AN}$) couplings are obtained: $|g_{Ae}\times
g^3_{AN}| \leq 5.5\times 10^{13}$ and $|g_{A\gamma} \times g^3_{AN}| \leq 4.6\times10^{11} \rm{GeV^{-1}}$ at $m_A \leq$ 1 MeV
(90\% c.l.).

\section{Test of electron stability}

A new limit on the stability of the electron for decay into a neutrino and a single monoenergetic photon $e \rightarrow \nu +
\gamma$ was obtained \cite{Ago_2015A}. This new bound, $\tau \geq 6.6\times10^{28}$ yr at 90\% C.L., is two orders of magnitude
better than the previous limit obtained with Borexino prototype CTF.

\section{ACKNOWLEDGMENTS}
The Borexino program is made possible by funding from INFN (Italy), NSF (USA), BMBF, DFG, and MPG (Germany), RFBR: Grants 15-02-
02117 and 14-22-03031, RFBR-ASPERA-13-02-92440 (Russia), RSF: Grant 16-12-10369 (Russia), and NCN Poland
(UMO-2012/06/M/ST2/00426). We acknowledge the generous support and hospitality of the Laboratori Nazionali del Gran Sasso (LNGS).


\nocite{*}
\bibliographystyle{aipnum-cp}%

\begin{thebibliography}{99}

\bibitem{Bel_2014}  G. Bellini et al. (Borexino Collaboration), Final results of Borexino phase-I on low-energy solar neutrino spectroscopy, Phys.
Rev. D 89, 112007 (2014)

\bibitem{Bac_2012} H. Back et al . (Borexino Collaboration), Borexino calibrations: hardware, methods, and results.  JINST 7 P10018 (2012)

\bibitem{Bel_2011} G. Bellini et al. (Borexino Collaboration), Precision measurement of the 7Be solar neutrino interaction rate in Borexino,
 Phys. Rev. Lett. 107, 141302 (2011)

\bibitem{Bel_2012} G. Bellini et al. (Borexino Collaboration), Absence of day-night asymmetry of 862 keV Be-7 solar neutrino rate in Borexino and
MSW oscillation parameters, Phys. Lett. B 707, 22 (2012)

\bibitem{Bel_2010} G. Bellini et al. (Borexino Collaboration), Measurement of the solar 8B neutrino rate with a LS target and 3 MeV energy
threshold in the Borexino detector,  Phys. Rev. D82:033006 (2010)

\bibitem{Bel_2012A} G. Bellini et al. (Borexino Collaboration), First evidence of pep solar neutrinos by direct detection in Borexino,  Phys. Rev.
Lett. 108, 051302 (2012)

\bibitem{Bel_2014A} G. Bellini et al. (Borexino Collaboration), Neutrinos from the primary proton-proton fusion process in the Sun. Nature,
512:383–386, (2014)

\bibitem{Arp_2008} C. Arpesella et al. (Borexino Collaboration), New results on solar neutrino fluxes from 192 days of Borexino data, Phys.
Rev. Lett. 101:091302, (2008)

\bibitem{Ago_2015} M. Agostini et al. (Borexino Collaboration), Spectroscopy of geo-neutrinos from 2056 days of Borexino data,  Phys. Rev. D
92, 031101 (2015)

\bibitem{Bel_2011A} G. Bellini et al. (Borexino Collaboration), Study of solar and other unknown anti-neutrino fluxes with Borexino at LNGS, Phys. Lett. B 696, 191 (2011)

\bibitem{Bel_2010A} G. Bellini et al. (Borexino Collaboration), New experimental limits on the Pauli forbidden transitions in 12C nuclei
obtained with 485 days Borexino data, Phys. Rev. C81:034317 (2010)

\bibitem{Bel_2012B} G. Bellini et al. (Borexino Collaboration), Search for Solar Axions Produced in p(d,3He)A Reaction with Borexino
Detector, Phys. Rev. D 85, 092003 (2012)

\bibitem{Bel_2013} G. Bellini et al. (Borexino Collaboration), New limits on heavy sterile neutrino mixing in 8B-decay obtained with the
Borexino detector, Phys. Rev. D 88, 072010 (2013)

\bibitem{Ago_2015A} M. Agostini et al. (Borexino Collaboration), Test of Electric Charge Conservation with Borexino, Phys. Rev. Lett. 115, 231802 (2015)

\end{thebibliography}

\end{document}